\documentclass[pra, 12pt, showpacs, showkeys, eqsecnum]{revtex4}
\usepackage{graphics}

\newcommand {\Tr} {{\mbox{Tr}}}

\begin{document}
\title{What is tested when experiments test \\ that quantum dynamics is linear}
\author{Thomas F. Jordan}
\email[email: ]{tjordan@d.umn.edu}
\affiliation{Physics Department, University of Minnesota, Duluth, Minnesota 55812}

\begin{abstract}
Experiments that look for nonlinear quantum dynamics test the fundamental premise of physics that one of two separate systems can influence the physical behavior of the other only if there is a force between them, an interaction that involves momentum and energy. The premise is tested because it is the assumption of a proof that quantum dynamics must be linear. Here variations of a familiar example are used to show how results of nonlinear dynamics in one system can depend on correlations with the other. Effects of one system on the other, influence without interaction between separate systems, not previously considered possible, would be expected with nonlinear quantum dynamics. Whether it is possible or not is subject to experimental tests together with the linearity of quantum dynamics. Concluding comments and questions consider directions our thinking might take in response to this surprising unprecedented situation.
\end{abstract}


\maketitle

\section{Introduction}\label{one}

If an experiment tests that quantum dynamics is linear, it also tests assumptions from which it can be proved that quantum dynamics is linear. In the classic proof of Wigner and Bargmann \cite{WignerGroupTheory,Bargmann64}, the key assumption is that the dynamics does not change absolute values of inner products of state vectors. This assumption is not interesting enough by itself for notice of its experimental tests. When experiments were done, first \cite{Shimony,ShullEtAl,GahlerEtAl} to test the nonlinear Schrodinger equation proposed by Bialynicki-Birula and Mycielski \cite{BirulaMycielski}, and then \cite{Bollinger,Chupp,Walsworth,Majumder} to test the nonlinear quantum mechanics proposed by Weinberg \cite{WeinbergNlqAnn,WeinbergNlqPrl},  there was little expression of concern that a discovery of nonlinear quantum dynamics would imply that the assumption of Wigner and Bargmann is not true.

Now there is a new proof \cite{me80,Whylinear}, from an assumption that would not be easy to let go. It proves the basic linearity of quantum dynamics that density matrices are mapped linearly to density matrices in an interval of time. This links with previously established results to complete a simple derivation of the linear Schrodinger equation for situations where it can be expected to apply. The assumption for the proof that maps of density matrices are linear is just that the dynamics does not depend on anything outside the system, but the system can be described as part of a larger composite system together with another separate system. In particular, the dynamics can not depend on correlations with another system or be changed by a measurement made on another system. A discovery of nonlinear quantum dynamics would imply this assumption is not true.

The surprising property of nonlinear quantum dynamics acting here is that one of two separate systems can influence the other without there being an interaction made by a force - without exchange of momentum and energy - between them. Influence without interaction can happen when there are correlations between the two systems, probabilities for values of quantities in one system that are not independent of probabilities for values of quantities in other system. Then making a measurement on one system can change the result of nonlinear dynamics in the other system.

Here we consider the simple case where each of the two systems is a spin 1/2, let the dynamics give no interaction between them, and use variations of a familiar example to illustrate how results of nonlinear dynamics in one system can depend on correlations with the other. The example was originally proposed to argue that nonlinear quantum dynamics allows signals faster than light \cite{GisinExample}. It is adapted here to describe the more broadly basic kind of effect of one system on another that the new proof says nonlinear quantum dynamics implies. Relativity is not considered here. The original example used an entangled state. It is modified here to require only classical correlations.

First, some background is provided in Sec.~II, and Sec.~III reviews how the dependence on correlations that the proof shows is necessary for nonlinear quantum dynamics can not happen in ordinary linear quantum mechanics. The nonlinear dynamics of the example is described in Sec.~IV. Dependence on correlations is shown by changes in results of the nonlinear dynamics between situations with no correlations in Sec.~V and classical correlations in Sec.~VI and when changes are made in other classical correlations in Sec.~VII. There is a signal when the classical correlations of Sec.~VII are changed to entanglement in Sec.~VIII. 

Influence without interaction between separate systems, not previously considered possible, would be expected with nonlinear quantum dynamics. Whether it is possible or not is subject to experimental tests together with the linearity of quantum dynamics. The concluding Sec.~IX contains comments and questions about directions our thinking might take in response to this surprising unprecedented situation.

\section{Background}\label{two}

It is a simple step from linear to nonlinear quantum dynamics: the Hamiltonian is allowed to depend on the state. A Hamiltonian that depends on the wave function gives a nonlinear Schrodinger equation. A Hamiltonian that depends on mean values gives nonlinear equations of motion for mean values \cite{Whylinear}.

It is assumed here that at each time everything is the same as in ordinary linear quantum mechanics, so the question is just about the dynamics, about how things change in time. Then the statement that the Hamiltonian depends on the state \cite{me64,me66} is a simple description of the difference between ordinary linear quantum mechanics and Weinberg's nonlinear quantum mechanics. Actually, it gives a bit more than Weinberg's theory; it includes some simple examples that Weinberg's theory can not describe \cite{me66}.

The result of quantum dynamics is the time dependence of mean values for Hermitian operators representing physical quantities. This includes the time dependence of probabilities, which are mean values for projection operators. The result is the same whether it is obtained from the Schrodinger picture or the Heisenberg picture. The linearity considered here is that the equations of motion for the mean values are linear in that the time derivative of each mean value is a function of mean values that depends on the state in a linear way \cite{me80,Whylinear}. It must be a linear function of mean values. It can not be a product of mean values, as it is in the equations of motion (\ref{Eqspin}) for the example of nonlinear dynamics we will consider. This linearity does not mean that the equations of motion for operators are linear; the equations of motion in the Heisenberg picture can be nonlinear. The time derivative of an operator may be a product of operators.

\section{The linear norm}\label{three}

It is all very simple in ordinary linear quantum mechanics. Consider two separate quantum systems $S$ and $R$. If the state of the system of $S$ and $R$ combined is represented by a  density matrix $\Pi$, the state of $S$ is represented by the density matrix
\begin{equation}
\label{rhofrompi}
\rho  = \Tr_R \Pi .
\end{equation}
A probability $\langle P\rangle $ for a proposition represented by a projection operator $P$ for $S$ is
\begin{eqnarray}
\label{mvA}
\langle P \rangle & = & \Tr_S \Tr_R [P\Pi ] \\ \nonumber
& = & \Tr_S [P\Tr_R [\Pi ]] \\ \nonumber
& = & \Tr_S [P\rho ].
\end{eqnarray}

We assume there is no interaction between $S$ and $R$. Then the dynamics for a time $t$ in the system of $S$ and $R$ combined makes changes described by a unitary operator $U_tV_t$ with $U_t$ a unitary operator for $S$ and $V_t$ a unitary operator for $R$. The probability $\langle P\rangle $ is changed during the time $t$ to
\begin{eqnarray}
\label{lt}
\langle P \rangle_t & = & \Tr_S \Tr_R [U_t^\dagger V_t^\dagger PU_tV_t \Pi ] \\ \nonumber
& = & \Tr_S [U_t^\dagger PU_t \Tr_R [V_t^\dagger V_t \Pi ]] \\ \nonumber
& = & \Tr_S [U_t^\dagger PU_t \Tr_R [\Pi ]] \\ \nonumber
& = & \Tr_S [U_t^\dagger PU_t \rho ].
\end{eqnarray}
It depends only on the dynamics in $S$ and the state of $S$. It does not depend on the dynamics in $R$, the state of $R$, or correlations between $S$ and $R$. 

This probability for a proposition for $S$ is not changed by a measurement made on $R$. Suppose a measurement on $R$ tests propositions represented by projection operators $E_k$ for $R$ such that
\begin{equation}
\label{Ekprop}
E_j E_k  = \delta_{jk}, \quad  \quad  \sum_k E_k = 1_R.
\end{equation}
The probability that the proposition represented by $E_k$ is true is
\begin{equation}
\label{Ekprob}
\langle E_k \rangle  = \Tr_S \Tr_R [E_k \Pi ].
\end{equation}
If the result of the measurement is that the proposition represented by $E_k$ is true, then after the measurement the state of the system of $S$ and $R$ combined is represented by the density matrix
\begin{equation}
\label{Ekdm}
\Pi_k  = \frac{1}{\Tr_S \Tr_R [E_k \Pi E_k]} E_k \Pi E_k.
\end{equation}
With this knowledge, the probability for the truth of the proposition represented by $P$ for $S$ is calculated to be
\begin{equation}
\label{EkandP}
\Tr_S \Tr_R [P\Pi_k ] = \frac{1}{\Tr_S \Tr_R [E_k \Pi E_k]} \Tr_S \Tr_R [PE_k \Pi E_k ]
\end{equation}
if the proposition represented by $E_k$ is true, and altogether, with all the possibilities being considered, the probability for the truth of the proposition represented by $P$ for $S$, the sum of the joint probabilities for $P$ and $E_k$, is
\begin{eqnarray}
\label{PprobEksum}
\sum_k\Tr_S \Tr_R [E_k \Pi ] \, \Tr_S \Tr_R [P\Pi_k ] & = & \sum_k \frac{\Tr_S \Tr_R [E_k \Pi ]}{\Tr_S \Tr_R [E_k \Pi E_k]} \Tr_S \Tr_R [PE_k \Pi E_k ]\\ \nonumber
& = & \Tr_S \Tr_R [P\Pi \sum_k E_k ]\\ \nonumber
& = & \Tr_S [P\Tr_R [\Pi ]] \\ \nonumber
& = & \Tr_S [P\rho ] .
\end{eqnarray}
A time $t$ between the measurement that tests the propositions for $R$ and the measurement that tests the proposition for $S$ can be included by changing $P$ to
\begin{equation}
\label{Patt}
U_t^\dagger V_t^\dagger PU_tV_t = U_t^\dagger PU_t 
\end{equation}
as in Eqs.(\ref{lt}).

There may be correlations that can be observed when measurements are made on both $S$ and $R$. A measurement made on $S$ alone can not show any effect of the dynamics in $R$, the state of $R$, or correlations between $S$ and $R$, and it can not detect that a measurement has been made on $R$.

\section{Nonlinear dynamics}\label{four}

Departures from this linear norm when there is nonlinear dynamics can be seen with a simple example. It is a representative case of Weinberg's nonlinear quantum mechanics \cite{WeinbergNlqAnn,WeinbergNlqPrl}. It is for a spin $1/2$ described by Pauli matrices $\Sigma_1 , \Sigma_2 , \Sigma_3 $. The Hamiltonian is
\begin{equation}
\label{Hspin}
H = \epsilon \langle \Sigma_3 \rangle \Sigma_3 .
\end{equation}
The equations of motion for the mean values $\langle \Sigma_1 \rangle $, $\langle \Sigma_2 \rangle $, $\langle \Sigma_3 \rangle $ for this Hamiltonian are the same as for any Hamiltonian that is a number times a spin matrix,
\begin{eqnarray}
\label{Eqspin}
\frac{d}{dt}\langle \Sigma_1 \rangle  & = & -2\epsilon \langle \Sigma_3 \rangle \langle \Sigma_2 \rangle \nonumber \\
\frac{d}{dt}\langle \Sigma_2 \rangle  & = & 2\epsilon \langle \Sigma_3 \rangle \langle \Sigma_1 \rangle \nonumber \\
\frac{d}{dt}\langle \Sigma_3 \rangle  & = & 0.
\end{eqnarray}
These equations of motion are nonlinear. What makes them nonlinear is that the Hamiltonian depends on the state. The matrix $\Sigma_3 $ in $H$ is multiplied by the mean value $\langle \Sigma_3 \rangle $ which depends on the state. The consequence is that the equations of motion describe precession with a frequency that depends on the state.

The solutions of these equations of motion are that $\langle \Sigma_3 \rangle $ is a constant and
\begin{eqnarray}
\langle \Sigma_1 \rangle (t) & = & \langle \Sigma_1 \rangle (0) 
cos(2\epsilon \langle \Sigma_3 \rangle t) - \langle \Sigma_2 \rangle (0) sin(2\epsilon \langle \Sigma_3 \rangle t) \nonumber \\
\langle \Sigma_2 \rangle (t) & = & \langle \Sigma_2 \rangle (0) cos(2\epsilon \langle \Sigma_3 \rangle t)+ \langle \Sigma_1 \rangle (0 )sin(2\epsilon \langle \Sigma_3 \rangle t).
\end{eqnarray}

We will consider three different sets of initial conditions for these equations of motion. Set (a) has two cases. They are for the two states that are represented by eigenvectors of $\Sigma_3 $. Either $\langle \Sigma_3 \rangle (0 )$ is $1$ or $\langle \Sigma_3 \rangle (0 )$ is $-$1, and $\langle \Sigma_1 \rangle (0 )$ and $\langle \Sigma_2 \rangle (0 )$ are zero. Then
\begin{equation}
\langle \Sigma_1 \rangle (t) = 0 = \langle \Sigma_2 \rangle (t).
\end{equation} 
The initial condition set (b) also has two cases. They are for the two states represented by eigenvectors of $(1/\sqrt{2})(\Sigma_1 + \Sigma_3 )$. Either $\langle \Sigma_1 \rangle (0 )$ and $\langle \Sigma_3 \rangle (0 )$ are both $1/\sqrt{2}$ or they are both $-1/\sqrt{2}$, and $\langle \Sigma_2 \rangle (0)$ is zero. Then
\begin{equation}
\label{s2t}
\langle \Sigma_2 \rangle (t) = \pm \frac{1}{\sqrt{2}}sin(\pm \sqrt{2}\epsilon t) = \frac{1}{\sqrt{2}}sin(\sqrt{2}\epsilon t).
\end{equation} 

The initial condition set (c) is for the state that is a mixture of the two states represented by eigenvectors of $(1/\sqrt{2})(\Sigma_1 + \Sigma_3 )$, with probability $p$ for the state where $\langle \Sigma_1 \rangle (0 )$ and $\langle \Sigma_3 \rangle (0 )$ are both $1/\sqrt{2}$ and probability $1-p$ for the state where $\langle \Sigma_1 \rangle (0 )$ and $\langle \Sigma_3 \rangle (0 )$ are both $-1/\sqrt{2}$. Then $\langle \Sigma_2 \rangle (0)$ is zero, $\langle \Sigma_1 \rangle (0 )$ and $\langle \Sigma_3 \rangle (0 )$ are both $(2p-1)/\sqrt{2}$, and
\begin{equation}
\label{sp2t}
\langle \Sigma_2 \rangle (t) = \frac{2p-1}{\sqrt{2}}sin(\sqrt{2}(2p-1)\epsilon t).
\end{equation}

\section{With no correlations}\label{five}

Consider, again, two separate quantum systems $S$ and $R$. Let $S$ be the spin $1/2$ of Sec.~IV, described by the Pauli matrices $\Sigma_1 , \Sigma_2 , \Sigma_3 $. It will contain the nonlinear dynamics described in Sec.~IV. Suppose the state of the system of $S$ and $R$ combined is represented by a density matrix
\begin{equation}
\label{stateRSno}
\Pi  = \rho \mu 
\end{equation}
with $\rho $ a density matrix for $S$ and $\mu $ a density matrix for $R$, so there are no correlations between $S$ and $R$. 

Suppose a measurement is made on $R$ that tests propositions represented by orthogonal projection operators $E_k$ for $R$, which Eqs.(\ref{Ekprop}) describe. 
If the result of the measurement is that the proposition represented by $E_k$ is true, then after the measurement the state of the system of $S$ and $R$ combined is represented by the density matrix
\begin{equation}
\label{Ekdmno}
\Pi_k  = \frac{1}{\Tr_S \Tr_R [E_k \Pi E_k]} E_k \Pi E_k = \rho \frac{1}{\Tr_R [E_k \mu E_k]} E_k \mu E_k.
\end{equation}
A probability $\langle P\rangle $ for a proposition represented by a projection operator $P$ for $S$ is
\begin{eqnarray}
\label{mvAno}
\langle P \rangle & = & \Tr_S \Tr_R [P\Pi ]  \\ \nonumber
& = & \Tr_S [P\rho \Tr_R [ \mu ]] \\ \nonumber
& = & \Tr_S [P\rho ] \\ \nonumber
& = & \Tr_S [P\rho \Tr_R [\frac{1}{\Tr_R [E_k \mu E_k]} E_k \mu E_k]] \\ \nonumber
& = & \Tr_S \Tr_R [P\Pi_k ] .
\end{eqnarray}
It is not changed by the measurement.

Let the density matrix for $S$ be
\begin{equation}
\label{st2RSclass}
\rho   = p|\nearrow \rangle \langle \nearrow |
 + (1-p)|\swarrow \rangle \langle \swarrow |
\end{equation}
where $|\nearrow \rangle $ and $|\swarrow \rangle $ are eigenvectors of $(1/\sqrt{2})(\Sigma_1 + \Sigma_3 )$ for the eigenvalues $1$ and $-1$, and $p$ is a number between $0$ and $1$. Then $\langle \Sigma_2 \rangle $ is zero, and $\langle \Sigma_1 \rangle $ and $\langle \Sigma_3 \rangle $ are both $(2p-1)/\sqrt{2}$. Let these be the initial conditions $\langle \Sigma_2 \rangle (0)$, $\langle \Sigma_1 \rangle (0 )$ and $\langle \Sigma_3 \rangle (0 )$ for the nonlinear dynamics. They are the initial conditions (c), so the result for $\langle \Sigma_2 \rangle (t)$ from the nonlinear dynamics is Eq.(\ref{sp2t}). This result is not changed by a measurement made on $R$.

\section{With classical correlations}\label{six}

Consider, still, two separate quantum systems $S$ and $R$, with $S$ the spin $1/2$ of Sec.~IV described by the Pauli matrices $\Sigma_1 , \Sigma_2 , \Sigma_3 $. Let $S$ contain the nonlinear dynamics described in Sec.~IV. There will be no other dynamics here. There will be no dynamics in $R$ and no interaction between $S$ and $R$. The dynamics will describe nothing interacting with $S$. 

Again, let $|\nearrow \rangle $ and $|\swarrow \rangle $ be eigenvectors of $(1/\sqrt{2})(\Sigma_1 + \Sigma_3 )$ for the eigenvalues $1$ and $-1$. Suppose the state of the system of $S$ and $R$ combined is represented by a density matrix
\begin{equation}
\label{stateRSclassp}
\overline{\Pi}  = p|\nearrow \rangle \langle \nearrow ||\alpha \rangle \langle \alpha |
 + (1-p)|\swarrow \rangle \langle \swarrow ||\beta \rangle \langle \beta |
\end{equation}
where $|\alpha \rangle$ and $|\beta \rangle$ are orthonormal vectors for $R$ and, again, $p$ is a number between $0$ and $1$. The probabilities calculated from this density matrix (\ref{stateRSclassp}) apply to an ensemble that can be prepared without involving either an interaction between the two systems or their quantum properties. Copies of system $S$ in the state described by $|\nearrow \rangle $ can be brought together without an interaction with copies of system $R$ in the state described by $|\alpha \rangle$ and mixed in the ratio $p/(p-1)$ with copies of $S$ in the state described by $|\swarrow \rangle $ brought together without an interaction with copies of $R$ in the state described by $|\beta \rangle$. 

Suppose a measurement is made on $R$ that tests the propositions represented by $|\alpha \rangle \langle \alpha |$ and $|\beta \rangle \langle \beta |$. After the measurement, the density matrix for the system of $S$ and $R$ combined is either $|\nearrow \rangle \langle \nearrow ||\alpha \rangle \langle \alpha |$ or $|\swarrow \rangle \langle \swarrow ||\beta \rangle \langle \beta |$ so $\langle \Sigma_1 \rangle $ and $\langle \Sigma_3 \rangle $ are either both $1/\sqrt{2}$ or both $-1/\sqrt{2}$. Let this be when t is zero and suppose that then the spin described by $\Sigma_1 , \Sigma_2 , \Sigma_3 $ is subject to the nonlinear dynamics described in Sec.~IV. The initial conditions (b) apply, so the result for $\langle \Sigma_2 \rangle (t)$ from the nonlinear dynamics is Eq.(\ref{s2t}). This is different from the result (\ref{sp2t}) obtained when there are no correlations.

The density matrix for $S$ is the same,
\begin{equation}
\rho  = p|\nearrow \rangle \langle \nearrow | \, + \, (1-p)|\swarrow \rangle \langle \swarrow |,
\end{equation}
whether the system of $S$ and $R$ combined is in the state with no correlations described by the density matrix (\ref{stateRSno}) or the state with correlations between $S$ and $R$ described by the density matrix (\ref{stateRSclassp}). Even though nothing is interacting with $S$, the results of the nonlinear dynamics in $S$ are not determined by the density matrix for $S$ and the dynamics in $S$. The density matrix $\rho $ for $S$ does not provide the information needed to establish the initial conditions for the nonlinear dynamics. The needed information involves the presence or absence of correlations between $S$ and $R$. It depends on the situation of $S$ in the larger system of $S$ and $R$ combined. The results of the nonlinear dynamics in $S$ involve the relation between $S$ and $R$ even though the dynamics includes no interaction between $S$ and $R$.

The measurement provides a way to reach a firm conclusion that results of nonlinear dynamics in $S$ can depend on correlations with $R$. To make sure that there is no room for doubt, we can consider only situations where there is time for a signal slower than light to bring information that the measurement has been made on $R$ before results of the dynamics in $S$ are observed. We can let the time when this information arrives be when $t$ is zero and the nonlinear dynamics in $S$ starts. Then it is certain that the nonlinear dynamics in $S$ will be for the initial conditions established by the state of the system of $S$ and $R$ that comes from the measurement on $R$.

\section{When correlations are changed}\label{seven}

Consider, yet again, two separate quantum systems $S$ and $R$, with $S$ the spin $1/2$ of Sec.~IV described by the Pauli matrices $\Sigma_1 , \Sigma_2 , \Sigma_3 $, and let $S$ contain the nonlinear dynamics described in Sec.~IV. Here again, there will be no other dynamics. There will be no dynamics in $R$ and no interaction between $S$ and $R$. The dynamics will describe nothing interacting with $S$. 

Let $|\uparrow \rangle $ and $|\downarrow \rangle $ be eigenvectors of $\Sigma_3 $ for the eigenvalues $1$ and $-1$. Suppose the state of the system of $S$ and $R$ combined is represented now by a density matrix
\begin{equation}
\label{stateRSclass}
\Pi  = \frac{1}{2}|\uparrow \rangle \langle \uparrow ||\alpha \rangle \langle \alpha |
 + \frac{1}{2}|\downarrow \rangle \langle \downarrow ||\beta \rangle \langle \beta |
\end{equation}
where $|\alpha \rangle$ and $|\beta \rangle$ are orthonormal vectors for $R$, and suppose a measurement is made on $R$ that tests the propositions represented by $|\alpha \rangle \langle \alpha |$ and $|\beta \rangle \langle \beta |$. After the measurement, the density matrix for the system of $S$ and $R$ combined is either $|\uparrow \rangle \langle \uparrow ||\alpha \rangle \langle \alpha |$ or $|\downarrow \rangle \langle \downarrow ||\beta \rangle \langle \beta |$ so $\langle \Sigma_3 \rangle $ is either $1$ or $-1$. Let this be when t is zero and suppose that then the spin described by $\Sigma_1 , \Sigma_2 , \Sigma_3 $ is subject to the nonlinear dynamics described in Sec.~IV. The initial conditions (a) apply, so $\langle \Sigma_2 \rangle (t)$ is zero.

Suppose that, instead, the state of the system of $S$ and $R$ combined is represented by a density matrix
\begin{equation}
\label{st2RSclass}
\overline{\Pi}  = \frac{1}{2}|\nearrow \rangle \langle \nearrow ||\alpha \rangle \langle \alpha |
 + \frac{1}{2}|\swarrow \rangle \langle \swarrow ||\beta \rangle \langle \beta |
\end{equation}
where, again, $|\nearrow \rangle $ and $|\swarrow \rangle $ are eigenvectors of $(1/\sqrt{2})(\Sigma_1 + \Sigma_3 )$ for the eigenvalues $1$ and $-1$ and $|\alpha \rangle$ and $|\beta \rangle$ are orthonormal vectors for $R$. Then, after a measurement is made on $R$ that tests the propositions represented by $|\alpha \rangle \langle \alpha |$ and $|\beta \rangle \langle \beta |$, the density matrix for the system of $S$ and $R$ combined comes out to be either $|\nearrow \rangle \langle \nearrow ||\alpha \rangle \langle \alpha |$ or $|\swarrow \rangle \langle \swarrow ||\beta \rangle \langle \beta |$, so $\langle \Sigma_1 \rangle $ and $\langle \Sigma_3 \rangle $ are either both $1/\sqrt{2}$ or both $-1/\sqrt{2}$. Now the initial conditions (b) apply, so now $\langle \Sigma_2 \rangle (t)$ is not zero; it is the oscillating solution of the equations of motion given by Eq.(\ref{s2t}). The results of the nonlinear dynamics in $S$ are changed when the density matrix for the system of $S$ and $R$ combined is changed from (\ref{stateRSclass}) to (\ref{st2RSclass}).

The density matrix for $S$ remains the same,
\begin{equation}
\frac{1}{2}|\uparrow \rangle \langle \uparrow  | \, +  \, \frac{1}{2}|\downarrow \rangle \langle \downarrow | = 
\frac{1}{2}|\nearrow \rangle \langle \nearrow | \, + \, \frac{1}{2}|\swarrow \rangle \langle \swarrow |,
\end{equation}
when the density matrix for the system of $S$ and $R$ combined is changed from (\ref{stateRSclass}) to (\ref{st2RSclass}). Even though nothing is interacting with $S$, the results of the nonlinear dynamics in $S$ are not determined by the density matrix for $S$ and the dynamics in $S$. Nonlinear dynamics can give different results for different mixtures that give the same density matrix \cite{HaagBannier}. As in Sec.~VI, the density matrix for $S$ does not provide the information needed to establish the initial conditions for the nonlinear dynamics in $S$. The needed information involves correlations between $S$ and $R$. It depends on the situation of $S$ in the larger system of $S$ and $R$ combined. The results of the nonlinear dynamics in $S$ involve the relation between $S$ and $R$ even though the dynamics includes no interaction between $S$ and $R$.

Again, the measurement provides a way to reach a firm conclusion that results of nonlinear dynamics in $S$ can depend on correlations with $R$. To make sure that there is no room for doubt, we can consider only situations where there is time for a signal slower than light to bring information that the measurement has been made on $R$ before the nonlinear dynamics in $S$ starts. Then it is certain that the initial conditions for the nonlinear dynamics in $S$ are specified by the state of the system of $S$ and $R$ that comes from the measurement on $R$.

\section{With entanglement}\label{eight}

Suppose $R$ is another spin $1/2$, like $S$. Let $|\ddot{\uparrow }\rangle $, $|\ddot{\downarrow }\rangle $ and $|\ddot{\nearrow }\rangle $, $|\ddot{\swarrow }\rangle $ be the state vectors for $R$ that are the analogs of $|\uparrow \rangle $, $|\downarrow \rangle $ and $|\nearrow \rangle $, $|\swarrow \rangle $ for $S$. Suppose the system of $S$ and $R$ combined is in the state where the total spin is zero. It is represented by
\begin{equation}
\frac{1}{\sqrt{2}}|\uparrow \rangle |\ddot{\downarrow }\rangle   \, -  \, \frac{1}{\sqrt{2}}|\downarrow \rangle |\ddot{\uparrow }\rangle \, = \,
\frac{1}{\sqrt{2}}|\nearrow \rangle |\ddot{\swarrow }\rangle \, - \, \frac{1}{\sqrt{2}}|\swarrow \rangle |\ddot{\nearrow }\rangle .
\end{equation}

As before, let $S$ contain the nonlinear dynamics described in Section IV, and assume there is no other dynamics. There is no dynamics in $R$, and no interaction between $S$ and $R$. The dynamics describes nothing interacting with $S$.

Suppose a measurement on $R$ tests the propositions represented by $|\ddot{\uparrow }\rangle \langle \ddot{\uparrow }|$ and $|\ddot{\downarrow }\rangle \langle \ddot{\downarrow }|$. Then after the measurement the density matrix for $S$ and $R$ is either $|\uparrow \rangle \langle \uparrow ||\ddot{\downarrow }\rangle \langle \ddot{\downarrow }|$ or $|\downarrow \rangle \langle \downarrow ||\ddot{\uparrow }\rangle \langle \ddot{\uparrow }|$ so $\langle \Sigma_3 \rangle $ is either $1$ or $-1$, the initial conditions (a) apply, and $\langle \Sigma_2 \rangle (t)$ is zero. If, instead, the measurement tests the propositions represented by $|\ddot{\nearrow }\rangle \langle \ddot{\nearrow }|$ and $|\ddot{\swarrow }\rangle \langle \ddot{\swarrow }|$, the density matrix for $S$ and $R$ comes out to be either $|\nearrow \rangle \langle \nearrow ||\ddot{\swarrow }\rangle \langle \ddot{\swarrow }|$ or $|\swarrow \rangle \langle \swarrow ||\ddot{\nearrow }\rangle \langle \ddot{\nearrow }|$, so $\langle \Sigma_1 \rangle $ and $\langle \Sigma_3 \rangle $ are either both $1/\sqrt{2}$ or both $-1/\sqrt{2}$, the initial conditions (b) apply, and $\langle \Sigma_2 \rangle (t)$ is the oscillating function (\ref{s2t}).

What is new here compared to Sec.~VII is that whether the initial conditions are (a) or (b) is determined just by the choice of the measurement made on $R$. This requires an entangled state for $S$ and $R$ from which the measurement can produce states with different correlations. In Sec.~VII there are only classical correlations and the different correlations are in states that are different from the start. 

The result here is a signal from $R$ to $S$. The signal is sent from $R$ when a measurement is chosen and made. It is read at $S$ when $\langle \Sigma_2 \rangle (t)$ is observed. This is how the example was originally proposed \cite{GisinExample}. The idea then was that the signal could be faster than light. It still does appear likely that nonlinear quantum dynamics would bring in signals faster than light \cite{me71,CzachorDoebner}, but it is not proved \cite{me64,me71,CzachorDoebner,Kent,Whylinear}.

Including classical correlations frames the issue more broadly. Signal or not, anything not in $S$ that changes the results of the dynamics in $S$ when there is nothing that interacts with $S$ would be a fundamental revision of physics.

\section{Comments and questions}\label{nine}

When one of two separate systems influences the physical behavior of the other, we say it is because there is a force between them. There is an interaction that involves momentum and energy. It is generally described by a Hamiltonian. Four kinds of fundamental force have been found. We would be surprised to find another. We might be even more surprised to find that there is another, completely different way that one of two separate systems can influence the physical behavior of the other. Until now, physics has not shown us how this could happen.

Now we see that with nonlinear quantum dynamics there would be influence without interaction between separate systems. There would be effects on the dynamics in one system caused by relations with a separate system even when there is no interaction between the two systems. Correlations would carry influences, effects of one system on the other. Both the new proof and the examples described here show that influence without interaction would be common. It would be a quite general result of nonlinear quantum dynamics in one system, measurements on a separate system, and correlations, but no interactions, between the two systems. No special contrivance is needed. It is all a direct and immediate consequence of the basic theory. The nonlinear quantum mechanics is a simple extension of ordinary linear quantum mechanics that can be easily described in the same language \cite{me64,me66}. Everything looks natural and reasonable.

Although it seems to be possible in theory, nonlinear quantum dynamics appears unlikely to be found in nature. The behavior needed to avoid signals faster than light looks unnatural and unreasonable \cite{me71}. Very accurate experiments that tested for nonlinear quantum dynamics did not find any. It may be more promising to look for nonlinear quantum dynamics at the extremes of nature, perhaps with theories of quantum gravity, where the relation of quantum dynamics to space-time and relativity may bend.

We can not say nonlinear quantum dynamics is impossible. It has not been shown to be logically inconsistent. There are consistent examples for individual systems. It has not been shown to necessarily predict, by itself, something experiments have found false. So we have to admit that physics might allow influence without interaction between separate systems. Nonlinear quantum dynamics requires that it is possible.

The assumption in the new proof that quantum dynamics must be linear is just that there is no influence without interaction between separate systems. This is an assumption most physicists want to make. It is far more compelling than the assumption in the classic proof of Wigner and Bargmann. But it is still an assumption. We are not sure that experiments will never show it is wrong. We can not order nature to follow assumptions we favor.

Do we believe the assumption? Should we test it? Do we need to test what we believe? Are experiments that look for nonlinear quantum dynamics less interesting now because there is little expectation of finding anything so far out of bounds of established physics? Or are they more interesting because they put those bounds to experimental tests?

Observation of nonlinear quantum dynamics in nature would change physics profoundly. Searches for nonlinear quantum dynamics that find none leave us to wonder about the physics we have. Is there a reason in principle, not yet found, that quantum dynamics absolutely must be linear? Or is the absence of large-scale nonlinear quantum dynamics a wonderful unnecessary property of nature for which we should be grateful, because it makes physics so much simpler?

\newpage

\bibliography{Nonlinear}
\end{document}